\documentclass[twocolumn, onecolappendix
]{aastex701}
\pdfoutput=1

\received{2025 September 26}
\revised{2025 December 23}
\accepted{2026 January 10}
\submitjournal{ApJ}

\graphicspath{{./}}
\usepackage{multirow}
\usepackage{amsmath}
\usepackage[T1]{fontenc}

\newcommand{\Lr}{L_{\rm R}}
\newcommand{\Lx}{L_{\rm X}}
\newcommand{\Lbol}{L_{\rm bol}}
\newcommand{\Ledd}{L_{\rm Edd}}
\newcommand{\ledd}{\ell_{\rm Edd}}

\newcommand{\Rx}{\mathcal{R}_{\rm X}}

\newcommand{\Ro}{\mathcal{R}_{\rm O}}
\newcommand{\mbh}{M_{\rm BH}}
\newcommand{\mdot}{\dot m}
\newcommand{\Mdot}{\dot M}
\newcommand{\msun}{M_\odot}
\newcommand{\ha}{H$\alpha$}

\newcommand{\aox}{\alpha_{\rm ox}}
\newcommand{\daox}{\Delta \alpha_{\rm ox}}
\newcommand{\fweak}{f_{\rm weak}}

\newcommand{\alphar}{\alpha_{\rm r}}
\newcommand{\lrlx}{L_{\rm R}/L_{\rm X}}
\newcommand{\fwhmhbeta}{{\rm FWHM}_{\rm H\beta}}
\newcommand{\Finc}{\mathcal{F}_{\rm inc}}

\shorttitle{Ruling out Compact Jets in Radio-quiet, High-Eddington AGNs}
\shortauthors{Paul \& Plotkin}

\begin{document}

\title{Ruling Out Compact Jets as the Dominant Source of Radio Emission\\ in Radio-quiet, High Eddington-ratio Active Galactic Nuclei}

\correspondingauthor{Jeremiah D. Paul} 

\author[0000-0003-0040-3910]{Jeremiah D. Paul}
\affiliation{Department of Physics \& Astronomy, Texas Tech University, Lubbock, TX, 79409, USA}
\affiliation{Department of Physics, University of Nevada, Reno, NV 89557, USA}
\email[show]{someagnguy@gmail.com / jerepaul@ttu.edu}

\author[0000-0002-7092-0326]{Richard M. Plotkin}
\affiliation{Department of Physics, University of Nevada, Reno, NV 89557, USA}
\affiliation{Nevada Center for Astrophysics, University of Nevada, Las Vegas, NV 89154, USA}
\email{rplotkin@unr.edu}

\begin{abstract}
The origin of core radio emission in radio-quiet active galactic nuclei (AGNs) is still actively debated. General relativistic magnetohydrodynamics simulations often predict the launching of moderately large-scale jets from super-Eddington accretion flows, but this prediction seems at odds with observations indicating most high/super-Eddington AGNs appear radio quiet. Here, we use the ratio of radio to X-ray luminosities as a multiwavelength diagnostic to probe the origin of radio emission in a sample of 69 radio-quiet, high/super-Eddington AGNs with black-hole masses $\mbh \sim 10^{5}$--$10^{9}~\msun$. With this wide dynamic range in $\mbh$, we adapt existing formalisms for how jetted radio emission and accretion-powered X-ray emission scale with black hole mass into the super-Eddington regime. We find that the radio/X-ray luminosity ratios observed across this $\mbh$ range are inconsistent with a jet-dominated model for radio emission. We discuss how our results may instead be consistent with a corona-dominated radio emission origin with a contribution from outflows at higher accretion rates.

\end{abstract}


\section{Introduction} \label{sec:intro3}

The central engine of an active galactic nucleus (AGN), powered by a luminous disk of gas accreting onto a galaxy's central black hole, produces radiation across the electromagnetic spectrum. The spectral character of this radiation varies by wavelength, dependent on the physical processes producing the emission. For example, the accretion disk emits blackbody-like radiation that peaks in the ultraviolet (UV), and X-rays are produced mainly via inverse Compton scattering of disk photons by a compact, hot corona. In addition to radiation, an AGN may also launch winds, jets, or other outflows that can reduce the energy and mass available for black hole growth \citep[e.g.,][]{Skadowski13, Giustini19}. All of these accretion products bear implications for the co-evolution of the black hole with the host galaxy, as well as its interaction with the intergalactic medium \citep[e.g.,][]{Fabian12, Kormendy13, King15, Yang21}.

Radio emission offers crucial opportunities to examine AGN outflows. Around 10\% of bright AGNs can be described as ``radio-loud'' via the radio--optical flux density ratio $\Ro=f\textsubscript{5 GHz}/f\textsubscript{4400 {\AA}} > 10$ (where $f\textsubscript{\rm 5 GHz}$ and $f\textsubscript{4400 {\AA}}$ are, respectively, the 5~GHz radio and 4400~{\AA} optical flux densities; e.g., \citealt{Kellermann89, Stocke92}). The radio emission from these objects is dominated by strong, relativistic jets that can extend to Mpc scales.

On the other hand, the remaining $\sim$\,90\% of AGNs appear ``radio-quiet'' ($\Ro < 10$) and typically do not show signs of extended jets, yet their outflows may still have an influence on their surroundings \citep[e.g.,][]{Girdhar22, Ulivi24}. Core radio emission in radio-quiet AGNs can originate from multiple processes with complex and varied spectral signatures and morphologies, as summarized below (see \citealt{Panessa19} for a review). At GHz frequencies, compact (i.e., unresolved) jets will be optically thick with a flat/inverted spectrum (spectral index $\alphar \gtrsim -0.5$, where the flux density goes as $S_{\nu} \propto \nu^{\alphar}$) from the overlap of multiple self-absorbed components (e.g., \citealt{Blandford79}), while extended or weak/failed jets will produce steep-spectrum ($\alphar \lesssim -0.5$), optically thin synchrotron emission. A compact corona will likely be unresolved, with a flat spectrum at GHz frequencies from concentric spherical components with power-law spectra that are self-absorbed at low energies \citep[e.g.,][]{Raginski16}. A disk wind may produce bremsstrahlung free-free radio emission with a flat spectrum ($\alphar \approx 0.1$), and wind shocks may produce synchrotron radio emission with a steep spectrum at GHz frequencies \citep[e.g.,][]{Nims15, Yamada24}. Star formation activity will also often show a steep spectrum at GHz frequencies. Because multiple sources can contribute simultaneously, the decomposition and diagnosis of radio emission pose a challenge, particularly for sources that are either too compact or too distant to resolve morphologically with today's radio observatories.

General relativistic magnetohydrodynamics simulations of super-Eddington accretion flows (Eddington ratios $\ledd > 1$)\footnote{We denote the Eddington ratio via $\ledd = L_{\rm bol}/L_{\rm Edd}$, where $L_{\rm bol}$ is the bolometric luminosity and $L_{\rm Edd} = 1.26\times 10^{38}(M/M_\odot)$~erg\,s$^{-1}$ is the Eddington luminosity.} tend to feature efficient, relativistic jets when configured with high black-hole spin and a magnetically arrested disk \citep[e.g.,][]{BlandfordZnajek, Narayan03, Tchekhovskoy11, Ricarte23b}. Intriguingly, this is somewhat at odds with observations, as jets are observed primarily (though not exclusively) from sub-Eddington AGNs ($\ledd \lesssim 10^{-2}$), while higher-Eddington AGNs have a greater tendency to be radio quiet \citep[e.g.,][]{Kellermann89, Greene06}. It may be that sufficient magnetic saturation and jet production occurs only sporadically in some AGNs, allowing them to appear radio quiet at particular epochs or on a time-averaged basis \citep[e.g.,][]{Wang25}. It has also been hypothesized that energy loss due to jet ejection can gradually cause a black hole to spin down \citep[e.g.,][]{Narayan22, Ricarte23b}, reducing jet efficiency and increasing black hole growth efficiency \citep{Massonneau23}. Alternatively, a low-power jet may be driven by a frozen-in and co-rotating disk magnetic field \citep{BlandfordPayne}, or by a combination of the above mechanisms \citep[e.g.,][]{Wang08, Zhang25}. Unfortunately, measurements of black hole spin and magnetic saturation pose a significant challenge with present resources, as they are best measured from extremely high-quality X-ray spectra \citep[e.g.,][]{Reynolds21}. 

Given this tension between theory and observation, improved constraints are needed to help describe how super-Eddington accretion disks may or may not launch jets. If not jets, what is the dominant source of radio emission in radio-quiet, super-Eddington AGNs? To help overcome the above challenges, we can take advantage of the fact that both radio and X-ray emission are expected to scale with black-hole mass and accretion rate \citep[e.g.,][]{Merloni02, Heinz03}. Indeed, observational studies have produced an empirical ``fundamental plane'' relationship wherein the radio luminosity in sub-Eddington AGNs is dependent on the X-ray luminosity (as a proxy for accretion rate) and black hole mass, extending down to stellar-mass black hole X-ray binaries (XRBs) in the low/hard state \citep[e.g.,][]{Merloni03, Falcke04}. Interpretation of this relationship typically invokes a jet dominating radio emission, so it is not certain whether a similar relationship extends meaningfully to radio-quiet AGNs at higher accretion rates \citep[e.g.,][]{Li08, Gultekin14, Bariuan22, Gultekin22}. Still, the formalisms developed for lower-Eddington objects are insightful and can be adapted for higher accretion rates to determine if radio emission is scaling as expected from a jet.

Herein, we employ the radio/X-ray luminosity ratio, $\Rx = \lrlx$, as a multiwavelength diagnostic to test for the presence of compact, unresolved jets in radio-quiet, high/super-Eddington AGNs. We gather a sample of 69 low-redshift ($z \lesssim 0.2$) AGNs with $\ledd \geq 0.1$ and a mass range $\mbh \sim 10^{5}$--$10^{9}~\msun$. The dynamic mass range of this sample allows us to examine radio jet scaling to lower masses in AGNs than ever before. Crucially, a major physical change is expected at $\ledd \gtrsim 0.3$: the accretion disk is thought to become radiatively inefficient, expanding at small radii and entering an advection-dominated ``slim-disk'' state \citep[e.g.,][]{Abramowicz88, Czerny19, Giustini19, Marziani25}. The luminosity of the inner disk will saturate, significantly reducing the observed $\ledd$ for a given accretion rate \citep[e.g.,][]{Mineshige00, Watarai00, Madau14}. It has also been suggested that the expanded disk and its outflows can shield the corona from our line of sight (dependent on orientation), leading some slim-disk AGNs to appear X-ray weak compared to their optical--UV output \citep[e.g.,][]{Wu12, Luo15, Ni18, Ni22} or their radio output \citep[e.g.,][]{Paul24}. Moreover, it may be that high disk luminosities lead to quenching of the corona \citep[e.g.,][]{Leighly07a, Leighly07b}, or that saturated production (i.e., a relative decrease) and/or shielding (by an outflow) of disk UV photons otherwise available for inverse-Compton upscattering contribute to reduced X-ray output \citep[e.g.,][]{Zappacosta20}. We therefore employ established models for the scaling of radio jets and X-ray emission with mass and accretion rate \citep{Heinz03, Markoff03, Merloni03}, but we adjust the formalisms involved to account for the saturated disk luminosities and the range of X-ray weaknesses expected in the slim-disk state. 

This paper is organized as follows. In Section \ref{sec:obs3} we describe the sample and data. In Section \ref{sec:models} we describe the models employed to diagnose jet radio emission. We present our results and discuss their implications along with alternative explanations in Section \ref{sec:discussion3}. Finally, we summarize our conclusions in Section \ref{sec:summ3}. We adopt the cosmological parameters $H_0=70$ km s$^{-1}$ Mpc$^{-1}$, $\Omega_{\rm M}=0.3$, and $\Omega_{\rm \Lambda}=0.7$. All uncertainties are reported at the $\pm$1$\sigma$ confidence level unless otherwise noted.

\section{Sample and Data} \label{sec:obs3}

\subsection{Sample Assembly} \label{subsec:samp3}

We desire a sample of high-Eddington AGNs across a wide range of masses (including lower masses than previously studied), with radio and X-ray coverage, and with rest-frame optical spectra covering H$\alpha$ and/or H$\beta$ (allowing us to re-estimate $\mbh$ and $\ledd$). For consistency in the radio, we began our search using samples from the literature containing optically selected, radio-quiet Type I AGNs with Karl G. Jansky Very Large Array (VLA) radio detections or upper limits obtained in either A or B configuration (for their relatively high spatial resolution). There are indications that 1.4~GHz emission is not a reliable tracer of nuclear activity in AGNs \citep[e.g.,][]{Gultekin14, Saikia18}, and, indeed, most studies of $\Rx$ have relied on frequencies $>$\,4~GHz. For consistency with many of these prior studies, and to avoid having to extrapolate from 1.4~GHz to higher frequencies along an uncertain (or assumed) spectral slope, we limited our search to detections and upper limits obtained at 5~GHz (VLA C-band), although in some cases we allowed observations at 10~GHz (VLA X-band; discussed below). We kept only radio-quiet objects with $\Ro \leq 10$. We also limited to $z \lesssim 0.2$ to help avoid evolutionary effects with redshift and ensure that the typical VLA synthesized beam sizes in A or B configuration translate to projected spatial radial extents of order $\lesssim 1$~kpc.

We then used the NASA/IPAC Extragalactic Database (NED) to search the literature for these samples' optical properties and X-ray coverage (if not already provided by the publication with radio data). X-ray observations were first gathered from the following resources (in order of preference, when multiple resources were available): Chandra X-ray Observatory (including the Chandra Source Catalog; \citealt{CSC24}), XMM-Newton, and the Swift X-ray Telescope. We required detections or upper limits within the 2--10~keV (hard) band to avoid possible soft X-ray excess from hotter accretion disk temperatures at lower $\mbh$ \citep[e.g.,][]{Haardt93, Done12} or host galaxy contamination (in observations at lower angular resolutions). If no X-ray coverage was available from a dedicated observing campaign with those resources, we crossmatched optical positions to the eROSITA all-sky survey (eRASS; \citealt{erass1, erass2}) using a $5\arcsec$ match radius. Typical positional uncertainties for objects detected in eRASS were $\sim$\,$1\arcsec$--$2\arcsec$. We checked for false positives by adding an offset of $10\arcsec$ to the declination coordinates and re-running the search (no false positives were found). If the object was in the eRASS footprint but no detection was found, we took the eRASS flux upper limit at the object's coordinates \citep{erasslim1}.

We describe the samples gathered below, listed in order of decreasing mass range. Going forward, we refer to these as ``subsamples'' using the given 3-character alphanumeric codes. 

\begin{itemize}
\setlength{\itemsep}{0pt}
    \item M93: The sample of \citet{Miller93} contains 90 quasars at $z \lesssim 0.5$ from the Palomar-Green (PG) bright quasar survey and presents re-reductions and re-measurements of VLA radio observations (originally from \citealt{Kellermann89}) taken in C-band (5~GHz) and A configuration. Requiring $\Ro \leq 10$ and $z \lesssim 0.2$ left 46 objects, and from these we found 31 objects with X-ray coverage. This subsample populates the high end of our mass range; $\mbh \approx 10^7$--$10^9 \msun$. 19 objects are detected at 5~GHz in A configuration; of these, 12 were classified as being consistent with a point source, and 7 showed some extended emission.\footnote{The \citet{Miller93} classification criteria included the ratio of A- to D-configuration peak flux densities at 5~GHz, rather than A-configuration peak-to-integrated values. As noted by \citet{Laor19}, the low-resolution D-configuration observations may include significant host contamination.} None were classified as showing a clear jet morphology.
    \item B18: \citet{Berton18} present VLA C-band (5~GHz), A-configuration radio observations of 74 narrow-line Seyfert 1 (NLS1) galaxies at $z \lesssim 1$. Requiring $\Ro \leq 10$ and $z \lesssim 0.2$ left 26 objects. We found that 8 of these objects overlapped with other subsamples (2 from M93, 5 from Y20, described below, and 1 from P24, also described below); after verifying that the reported radio flux densities are consistent between samples, we excluded these 8 objects from the B18 sample. From the 18 remaining objects we found 4 with X-ray coverage. Note that a number of radio sources in the full \citet{Berton18} sample were found to have moderate to high levels of diffuse extended emission. For the 4 in our subsample, two show minor, diffuse resolved emission (with possible, though unconfirmed, indications of an outflow morphology) while the other two remain compact, and all have peak-to-integrated flux density ratios (ratio$_{\rm p/i}$) $\gtrsim 90\%$. The mass range of this subsample is $\mbh \approx 10^6$--$10^8 \msun$.
    \item Y20: From a parent sample of 60 high Eddington-ratio NLS1s, \citet{Yang20} compile a sample of 24 super-Eddington accreting massive black holes (SEAMBHs; e.g., \citealt{Wang13, Wang14b, Du14}) at $z \lesssim 0.2$ with VLA C-band (5~GHz), A-configuration radio observations and 2--10~keV X-ray observations from a variety of literature resources. We excluded 5 objects already present in the M93 subsample and 1 object present in the P24 subsample, as well as 1 object with $\Ro > 10$, leaving us 17 SEAMBHs from this sample, with a mass range of $\mbh \approx 10^6$--$10^7 \msun$. All but two of these objects are consistent with a radio point source (ratio$_{\rm p/i}$ $\gtrsim 80\%$). One object, Ark 564, shows an elongated structure suggesting a mildly relativistic outflow .
    \item P24: \citet{Paul24} present a radio/X-ray analysis of 18 objects from the \citet{GH07s} parent sample of 174 high-confidence, low-mass AGNs. VLA radio data include X-band (10~GHz) observations taken in both A and B configurations. All 18 objects are at $z < 0.15$ and have Chandra X-ray observations. Requiring $\Ro \leq 10$ left us 15 objects from this sample, populating the low end of our mass range; $\mbh \approx 10^{5.5}$--$10^6 \msun$. 12 are radio-detected, and of these, 7 are point-like (ratio$_{\rm p/i}$ $\gtrsim 80\%$), 3 show minor extended emission, 1 shows primarily diffuse extended emission, and 1 shows a compact core component with significant diffuse extended emission that may be consistent with host morphology. None show clear jet signatures.
    \item G22: \citet{Gultekin22} present VLA C-band (5~GHz), A-configuration radio observations of 8 candidate massive black holes ($\mbh \lesssim 10^5 \msun$) from the sample of \citet{Reines13}. The objects in the parent sample were initially identified as candidate AGNs at $z \lesssim 0.05$ via analysis of their narrow emission line ratios on the BPT diagram (\citealt{BPT}; see also \citealt{Kewley06}), with these 8 objects being given further confidence via Chandra X-ray detections from \citet{Baldassare17}. Only 3 of the 8 objects with X-ray coverage met the $\Ro \leq 10$ requirement. Due to this small subsample size and the uncertainty in both their AGN classification and their mass estimates, we do not include these 3 remaining candidates in any of the following statistical analyses (though we do plot them in figures for visual comparison).
\end{itemize}

The above selection left 71 objects. We re-estimated $\mbh$ and $\ledd$ for all using the scheme described in Section \ref{subsec:data_mass_edd} and retained only those with $\ledd \geq 0.1$. This last cut left 69 radio-quiet, high Eddington-ratio AGNs in our final sample. 

\subsection{Sample Properties} \label{subsec:samp_prop3}

\begin{figure*}[t]
\epsscale{1.15}
\plotone{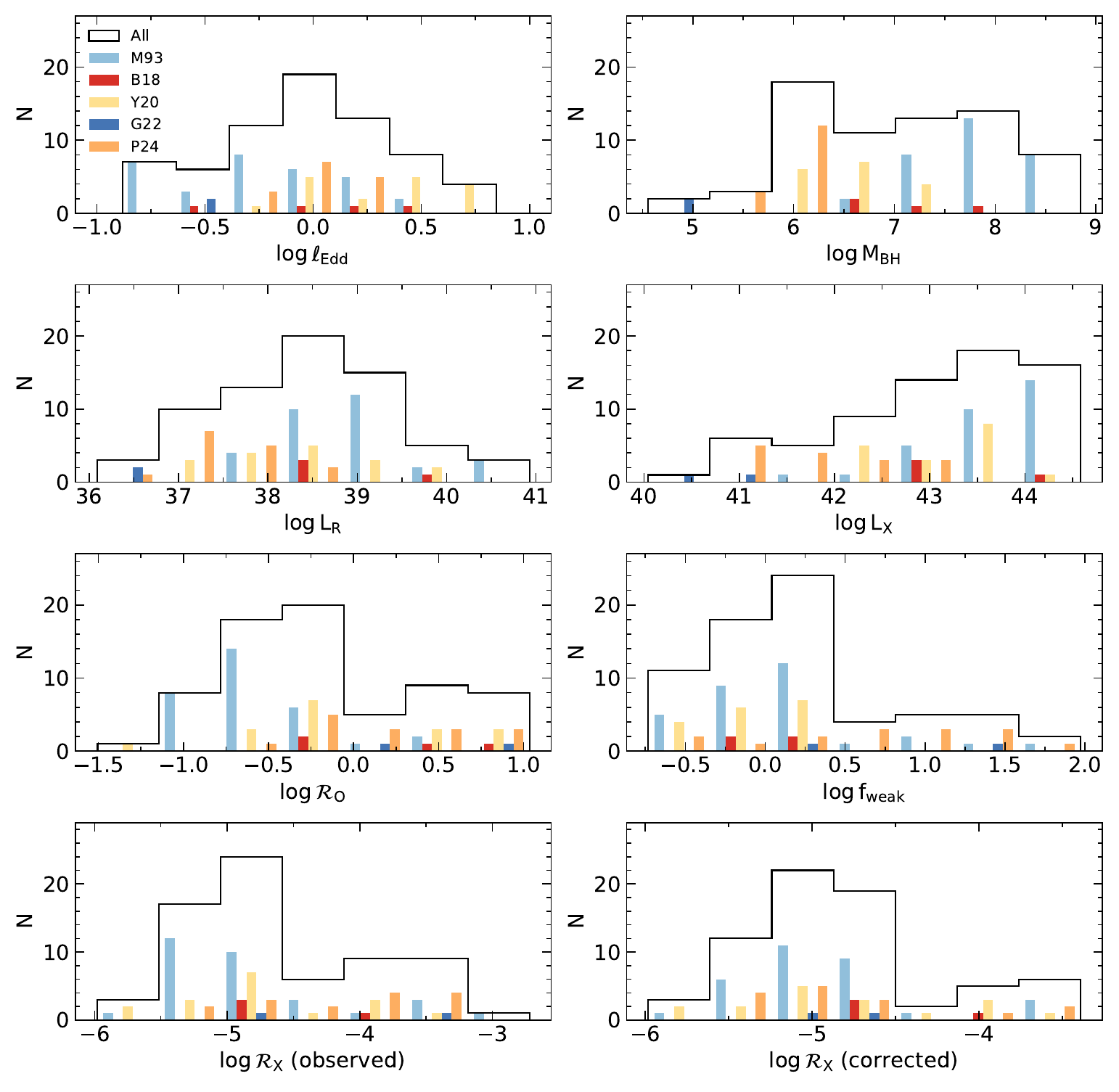}
\caption{Histograms showing the distributions of key sample properties. The full sample is shown by the black outline, and within each bin the subsample distributions are shown by colored bars (note that placement within each bin is based simply on sample order and does not necessarily reflect a fine x-axis value). In the lower-left panel we show the observed values of $\Rx$, while in the lower-right panel we show $\Rx$ corrected for X-ray weakness (where applicable; see Section \ref{subsec:data_lum}).
\label{fig:p3hist}
}
\end{figure*}

The distributions of our sample's key properties are illustrated in Figure \ref{fig:p3hist}. The optical properties of the sample are listed in Table \ref{tab:P3sampOPT}, and the radio and X-ray properties are listed in Table \ref{tab:P3sampRX} (see Appendix \ref{sec:app_tables}). 

We performed a number of tests to examine our sample for bias or other deficiencies. Via Kendall's $\tau$ tests, we found an apparent correlation between $\mbh$ and $z$ ($p_{\rm null} \sim 10^{-7}$) in our sample, but this is clearly attributable to a selection effect (i.e., lower-mass AGNs are more easily detected at lower $z$), and we did not find correlations between $z$ and $\ledd$ ($p_{\rm null} \sim 1$), $z$ and $\Rx$ ($p_{\rm null} \sim 0.1$), or $\mbh$ and $\Rx$ ($p_{\rm null} \sim 0.1$). We also found via a partial Kendall's $\tau$ test that, as expected, $\Lr$ and $\Lx$ remain highly correlated with each other even when corrected for a possible codependence on $z$ \citep[e.g.,][]{Singal19}, at a significance $p_{\rm null} \sim 10^{-7}$. Accordingly, we do not expect significant evolutionary effects to influence our results over our range of $z$.

We note that a small gap appears in the distribution of $\log \Rx$ at $\sim -4.25$ (visible in figures that follow), but we stress that this is likely a simple lack of coverage rather than a bimodal distribution. Not unexpectedly, the objects above this gap (at higher $\Rx$) do have somewhat higher values of the radio--optical ratio on average ($\langle \Ro \rangle \approx 5.7$) compared to the rest of the sample ($\langle \Ro \rangle \approx 1.0$), though they still qualify as radio-quiet. We discuss a possible relationship between $\Rx$ and $\ledd$ in Section \ref{subsec:p3_corona}, but the properties of our highest-$\Rx$ objects are otherwise unbiased (including sample, mass range, redshift, and X-ray weakness). For completeness, we performed a one-way analysis of variance (ANOVA, which tests the null hypothesis that two or more groups have the same population mean) and found that our subsamples are consistent in $\Rx$ at the $p_{\rm null} \approx 0.3$ level. We conclude that this gap could plausibly be filled by lower limits (from X-ray nondetections) or uncertainty on $\Rx$, or by additional observations. 

Acknowledging the above known biases and sample incompleteness, we believe that we have formed the most complete sample available with current data. Since these deficiencies are minor and understood, this sample allows us to probe relatively rare physics of super-Eddington accretion in a way that will inform building more complete samples for future studies at both low and high redshift.

\subsection{Recalculation of Optical, Radio, and X-ray Luminosities} \label{subsec:data_lum}

For the optical (where possible), radio, and X-ray wavebands we took fluxes from the literature to recalculate luminosities and avoid the issue of differing choices in cosmology. In the optical--UV, we found the continuum luminosities $L_{\rm 5100}$ and $l_{2500}$ using the following.\footnote{$L_{\rm 5100} = \nu L_{\nu}$ at 5100~{\AA} (in erg s$^{-1}$), and $l_{2500}$ is the monochromatic luminosity at 2500~{\AA} (in erg s$^{-1}$ Hz$^{-1}$).}  In general, we assumed a power-law continuum following the form $f_{\nu} \propto \nu^{-0.44}$ \citep[][]{VB01}. For the M93 subsample, we converted from 4681~{\AA} continuum luminosities reported by \citet{Davis11}. For B18 and Y20, 5100~{\AA} fluxes or luminosities were already available. For G22 and P24, we used \ha\ line luminosities ($L_{\rm H\alpha}$) along with the updated $L_{\rm H\alpha}$--$L_{\rm 5100}$ scaling relation from \citet[][see Eq.~2 and Table 5 therein]{Cho23}. References for the original optical data are given in Table \ref{tab:P3sampOPT}.

In radio, we found the radio luminosity $\Lr = \nu L_{\nu}$ at 5~GHz from the 5~GHz peak flux densities, except for the P24 subsample, which we converted from 10~GHz to 5~GHz assuming a power-law continuum of the form $S_{\nu} \propto \nu^{\alphar}$. Following the analysis performed by \citet{Paul24}, for this conversion we used measured in-band spectral indices only where the associated 1$\sigma$ errors are less than $\pm 0.4$, else we assumed $\alphar = -0.7$. 

In X-rays we found the 2~keV monochromatic luminosity $l_{\rm 2 keV}$ and the 2--10~kev luminosity $\Lx$ by extrapolating observed fluxes from the original band (e.g., typically 2--7~keV for Chandra, 2--5~keV for eRASS) to unabsorbed 2--10~keV fluxes using the Portable, Interactive Multi-Mission Simulator (PIMMS). We adopted Galactic column densities ($N_{\rm H,gal}$) from the HI4PI survey \citep{HI4PI}, and we utilized measured X-ray photon indices ($\Gamma$) when reported, else we assumed $\Gamma = 2.0 \pm 0.2$ as is considered typical of high-Eddington AGNs. 

A number of AGNs in our sample appear to be X-ray weak compared to their optical--UV luminosities, per the following. We began with the X-ray--UV broadband spectral index, 
$\aox = 0.38\log(l_{\rm 2 keV} / l_{\rm 2500})$ \citep[][]{Tananbaum79}. We adopted the customary X-ray deviation parameter,
$\daox = \aox - \alpha_{\rm ox,qso}$,
where $\alpha_{\rm ox,qso}$ is the value predicted by the $\aox$--$l_{\rm 2500}$ relationship displayed by broad-line quasars.\footnote{For consistency with prior works, we adopt the best-fit $\alpha_{\rm ox,qso}$ relationship given by Eq.~(3) of \citet{Just07}. \citet{Timlin20} suggest an intrinsic scatter of $\pm 0.11$~dex.} From the above, the X-ray weakness factor is
\begin{equation} \label{eq3}
\fweak = 10^{- \daox / 0.38},
\end{equation}
and we have defined ``X-ray-weak'' objects as having $\fweak \geq 6$.

Most of our X-ray-weak objects are from the P24 subsample. \citet{Paul24} argued that these objects may be affected by orientation-dependent X-ray shielding by the inner regions of a slim accretion disk. If we assume this to be true, we can correct for their X-ray weakness and obtain a simple estimate of their unshielded, intrinsic X-ray luminosities by applying $\fweak$: 
\begin{equation} \label{eq:Lxcorr}
L_{\rm X, corr} = \fweak \Lx.
\end{equation}
There are several additional X-ray-weak objects within the other subsamples for which we currently lack evidence beyond $\ledd$ to assume a slim disk. Nevertheless, if they are accreting at near/super-Eddington rates, there is the possibility they are being shielded. We therefore tested three different X-ray weakness correction schemes for our analyses: correct none for X-ray weakness, correct objects with $\fweak \geq 6$ in only the P24 subsample, and correct all objects with $\fweak \geq 6$. We have taken the last to be our fiducial version. In Table \ref{tab:P3sampRX} we provide the uncorrected $\Lx$ for all objects. Per the rule defined above, for those with $\fweak < 6$ we tabulate $\Rx = \lrlx$, and for those with $\fweak \geq 6$ we tabulate $\Rx = \Lr/L_{\rm X, corr}$.

\subsection{Recalculation of Masses, Eddington Ratios, and Accretion Rates} \label{subsec:data_mass_edd}

For consistency, we revised $\mbh$ and $\ledd$ estimates for all objects in our sample using the following recipes:

$\mbh$: For subsamples with broad H$\beta$ FWHM ($\fwhmhbeta$) measurements in the literature (M93, B18, Y20), we found the single-epoch virial mass estimate via $\mbh = f_{\rm BELR} R_{\rm BELR} V^2 / G$, where G is the gravitational constant, $f_{\rm BELR}$ is a scaling factor for the unknown geometry of the BELR, $R_{\rm BELR}$ is the BELR size, and $V$ is usually measured either from the H$\beta$ line dispersion ($\sigma_{\rm H\beta}$) or from $\fwhmhbeta$. We used the H$\beta$ BELR size-luminosity relationship of \citet{Woo24} to find $R_{\rm BELR}$ from $L_{\rm 5100}$ (see their Eq.~7 and Case 1 in their Table 7). We adopted $f_{\rm BELR} = 4.82$ as found by \citealt{Batiste17}, who used $V = \sigma_{\rm H\beta}$ in their derivation.\footnote{When using $V = \fwhmhbeta$, typically $f_{\rm BELR} \approx 1$ \citep[e.g.,][]{Onken04, Woo15}. The difference between this and our adopted method is negligible, particularly in comparison to our error budget (Section \ref{subsec:data_err}).} We can approximate $\sigma_{\rm H\beta}$ as $\fwhmhbeta/2$ \citep[e.g.,][and references therein]{YQChen22}, so we have used $V = \fwhmhbeta / 2$. Reliable $\fwhmhbeta$ measurements are not available for the low-mass subsamples (G22, P24; see, e.g., \citealt{Greene05}), so for these we instead used the broad H$\alpha$ line luminosity and FWHM with the conversion given by Eq.~(6) of \citet{Cho23}.

$\ledd$: For all objects, we found the Eddington luminosity as $\Ledd = 1.26 \times 10^{38} (\mbh / \msun)$~erg~s$^{-1}$. We used $L_{\rm 5100}$ with established bolometric corrections to find $L_{\rm bol}$. \citet{Netzer19} provides a bolometric correction of the form $L_{\rm bol} = 40 (L_{\rm 5100}/10^{42})^{-0.2}$ but cautions that the correction is derived assuming $\mbh \geq 10^7 \msun$ and $\ledd \leq 0.5$. The method of \citet{Netzer19} appears generally consistent with prior estimates for the higher-mass objects comprising the M93 subsample. However, this method severely underestimates the SEAMBH subsamples (B18 and Y20). Bolometric corrections for SEAMBHs (including NLS1s) tend to run much higher, with a range $K_{\rm bol} \sim 40$--150, where $L_{\rm bol} = K_{\rm bol} L_{\rm 5100}$ \citep[e.g.,][]{Wang13, Castello-Mor16}. Furthermore, the objects in P24 share similarities with super-Eddington NLS1s, meaning that for the P24 and G22 subsamples, the SEAMBH correction may be more appropriate than that of \citet{Netzer19}. We therefore tested three different bolometric correction schemes: correct all objects using \citet{Netzer19}, correct all using the SEAMBH factor, and correct using a sample-dependent mix (\citealt{Netzer19} for M93 and the SEAMBH factor for B18, Y20, G22, and P24). We used the latter (sample-dependent) version for the fiducial $\ledd$ values reported in Table \ref{tab:P3sampOPT}.\footnote{When applying the SEAMBH correction, we assumed $K_{\rm bol} = 40$, meaning our reported $\ledd$ may still constitute a lower estimate for those objects.} We also combined these three schemes with the three X-ray correction schemes described in Section \ref{subsec:data_lum}, resulting in a matrix of 9 total ``correction'' scenarios that allowed us to examine the influence of our assumptions. We found that across these 9 scenarios, our final results did not change substantially within the range of error. For brevity, we report only the results of the fiducial corrections (i.e., $\Rx$ corrections for all X-ray-weak objects via $\fweak$ and sample-dependent bolometric corrections). 

\begin{figure}[t]
\epsscale{1.15}
\plotone{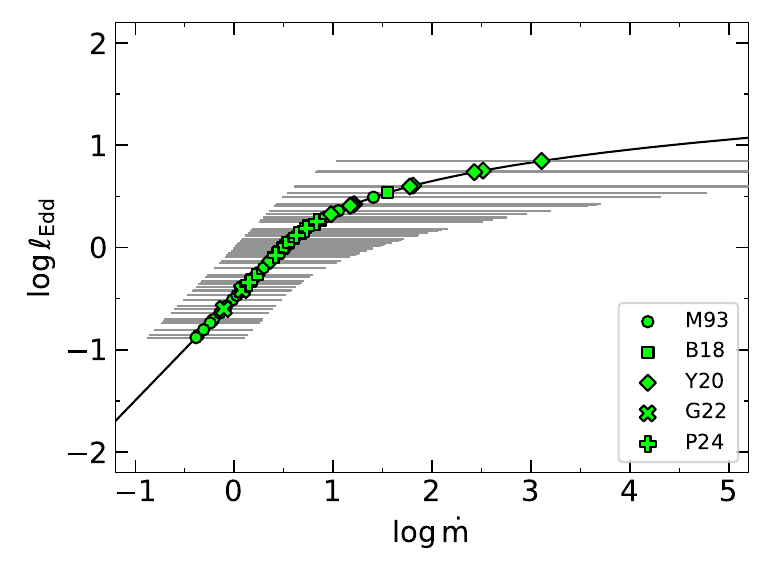}
\caption{Comparison of $\ledd$ to $\mdot$ as found using Eq.~(\ref{eq:mdot_saturated}). Our sample is shown by symbols given in the legend (categorized by subsample; Section \ref{subsec:samp3}). The grey bars illustrate error on $\mdot$ from a $\pm$0.5~dex variance in $\ledd$. The black curve shows the relationship given by Eq.~(8) of \citet{Mineshige00}, modified to use $\ledd = \Finc L/\Ledd$, where $L/\Ledd$ is the intrinsic Eddington ratio and $\Finc$ is our adopted inclination factor (see Eq.~\ref{eq:inc_fact}).
\label{fig:ledd_vs_mdot}
}
\end{figure}

We also require an estimate of the dimensionless Eddington accretion ratio, $\mdot \equiv \Mdot/\Mdot_{\rm Edd} = \eta \Mdot c^2/\Ledd$ (where $\Mdot$ is the mass inflow rate and $\eta$ is the mass-to-luminosity conversion efficiency) for the radio and X-ray emission models we discuss later. $\ledd$ is sometimes used as an idealized surrogate for $\mdot$ under radiatively efficient, thin-disk assumptions, but because the accretion disk luminosity is expected to saturate at high Eddington ratios in the slim-disk scenario, we cannot assume $\mdot \approx \ledd$. To address this, we invert Eq.~(8) of \citet{Mineshige00} to find:\footnote{Note that \citet{Mineshige00} set $\mdot  = {\dot M} c^{2} / \Ledd$, excluding $\eta$. We have reintroduced $\eta$ out of preference. This does not cause any issues with modeling (all sources are scaled by the same factor), but may necessitate care if attempting to compare $\mdot$ values with some other works. For this reason, we do not tabulate $\mdot$.}
\begin{equation} \label{eq:mdot_saturated}
\mdot \approx
\begin{cases}
25 \eta\ \frac{\ledd}{\Finc} & : \frac{\ledd}{\Finc} < 2,\\
50 \eta\ \exp{\left(\frac{1}{2} \frac{\ledd}{\Finc} - 1\right)} & : \frac{\ledd}{\Finc} \geq 2.
\end{cases}
\end{equation}
$\Finc$ is an inclination factor accounting for the non-isotropic nature of the disk radiation (see, e.g., Section 5 of \citealt{Mineshige00} and Section 2.1 of \citealt{Netzer19}):
\begin{equation} \label{eq:inc_fact}
\Finc\ =\ \frac{L_{\nu}}{L_{\nu}\textrm{(face-on)}}\ =\ \frac{1}{1+b} \cos{\theta}\ (1+b\cos{\theta}), 
\end{equation}
where $\theta$ is the inclination angle and $b$ accounts for limb-darkening. Following \citet{Netzer19}, we adopt a generalized $\Finc = 0.5$ (assuming a mean $\theta \approx 45$~deg and $b\approx2$) and $\eta = 1/16$. Figure \ref{fig:ledd_vs_mdot} illustrates this modeled relationship between $\ledd$ and $\mdot$ at high accretion rates. Note that once $\ledd$ begins to saturate, an error of 0.5~dex in $\ledd$ can translate to large (orders of magnitude) error in $\mdot$. The few objects in our sample at $\log \mdot \gtrsim 2$ are therefore highly uncertain, although we stress that we can be confident they are still very high-$\mdot$/super-Eddington.

\subsection{Error Analysis} \label{subsec:data_err}

In the analyses that follow, we account for error and nondetections using a Monte Carlo scheme to generate 2000 iterations of mock $\Lr$, $\Lx$, $\mbh$, and $\ledd$ measurements for each object. For radio detections, we randomly vary the flux density via a normal distribution, with the mean and standard deviation set by the observed flux density and error (if error was not originally reported, we adopt the background rms as the error), and we vary the spectral index $\alphar$ via a uniform distribution between $-0.7$ and 0.1 before recalculating $\Lr$. For X-ray detections, we randomly vary the flux and $\Gamma$ via a normal distribution, with the mean and standard deviation set by the measurement and error before recalculating $\Lx$ (if error was not originally reported, we adopt an error of 10\% of the measurement). We take the 16th and 84th percentile values of the resulting distributions as the $\pm 1\sigma$ error for each object. For radio or X-ray nondetections, flux values are instead drawn randomly from a uniform distribution between the reported upper limit and 50\% of the upper limit. We vary $\mbh$ and $\ledd$ via a uniform distribution with a $\pm 0.5$~dex assumed systemic error. For statistical tests and ordinary-least-squares fits to our sample, we test each Monte Carlo iteration and take our result from the mean and $\pm 1\sigma$ error (found from the 16th and 84th percentile values) of the distributions of those tests.

\section{Model to Identify Jets via \texorpdfstring{$\lrlx$}{the Radio/X-ray Luminosity Ratio}} \label{sec:models}

Following, e.g., \citet{Heinz03}, \citet{Markoff03}, \citet{Merloni03}, etc., we take the radio luminosity of the jet to be non-linearly dependent on the black hole mass and accretion rate: 
\begin{equation} \label{eq:FP_LR}
\Lr = K_{\rm r} \mbh^{\xi_{\rm M}} \mdot^{\xi_{\mdot}},
\end{equation}
where $\xi_{\rm M}$ and $\xi_{\mdot}$ are scaling powers and $K_{\rm r}$ is a normalization constant. The coronal X-ray luminosity is taken to be dependent linearly on mass but non-linearly on accretion rate:
\begin{equation} \label{eq:FP_LX}
\Lx = K_{\rm x} \mbh \mdot^{q},
\end{equation}
where $q$ is a (possibly $\mdot$-dependent) scaling power and $K_{\rm x}$ is a normalization constant. However, in the super-Eddington case we must also account for diminished $\Lx$ due to weakly coupled coronal output (see Section \ref{sec:intro3}). Using Eq.~(8) of \citet{Mineshige00} to define saturated and non-saturated bolometric luminosities ($L_{\rm bol,s}$ and $L_{\rm bol,n}$, respectively, for slim and thin disks), we can write the super-Eddington case of Eq.~(\ref{eq:FP_LX}) as $\Lx(\mdot/\eta \geq 50) = \Lx(\mdot/\eta < 50) L_{\rm bol,s}/ L_{\rm bol,n}$, or: 
\begin{equation} \label{eq:Lx_saturated_gen}
\Lx =
\begin{cases}
K_{\rm x} \mbh \mdot^{q} & : \frac{\mdot}{\eta} < 50,\\
50\eta K_{\rm x} \mbh \mdot^{(q-1)} A(\mdot)  & : \frac{\mdot}{\eta} \geq 50,
\end{cases}
\end{equation}
where $A(\mdot) = 1 + \ln(\frac{\mdot}{50 \eta})$. See Appendix \ref{sec:app} for additional validation of this approach. 

Taking the ratio of Eqs.~(\ref{eq:FP_LR}) and (\ref{eq:Lx_saturated_gen}), we find our model for $\Rx$:
\begin{equation} \label{eq:FP_LRLX}
\Rx = 
\begin{cases}
K_{\rm rx} \mbh^{(\xi_{\rm M}-1)} \mdot^{(\xi_{\mdot}-q)} & : \frac{\mdot}{\eta} < 50,\\
\frac{K_{\rm rx}}{50\eta} \mbh^{(\xi_{\rm M}-1)} \mdot^{(\xi_{\mdot}-q+1)} A(\mdot)^{-1} & : \frac{\mdot}{\eta} \geq 50,
\end{cases}
\end{equation}
where $K_{\rm rx} = K_{\rm r}/K_{\rm x}$. 

To find $\xi_{\rm M}$, $\xi_{\mdot}$, and $q$ for our model, we adopt the assumptions of \citet{Heinz03}: the jet emission is synchrotron radiation from a power-law electron distribution of the form ${\rm d}n/{\rm d}\gamma = C\gamma^{-p}$ (where $\gamma$ is the Lorentz factor of the particles, $C$ is a normalization constant, and $p$ is the power-law index, which we fix at $p = 2$), and the injection of the power-law electron distribution occurs at some fraction of equipartition with the magnetic field pressure ($C \propto B^2$). For the high Eddington ratios ($\ledd \geq 0.1$) found in our sample, radiation pressure is expected to dominate over gas pressure in the disk at small radii, so we take $\xi_{\rm M} = 17/12 + \alphar/3$, $\xi_{\mdot} = 0$, and $q=0.5$ \citep{Heinz03, Merloni03}.\footnote{Note that \citet{Heinz03} define $\xi_{\rm M} = 17/12 - \alphar/3$ via a sign difference in $\alphar$; i.e., $S_{\nu} \propto \nu^{-\alphar}$.} Assuming a compact, optically thick jet, we adopt a spectral index $\alphar=0$. Finally, we find the intercept $K_{\rm rx}$ by inverting Eq.~(\ref{eq:FP_LRLX}) and normalizing to the mean values of our sample, $\langle \log \Rx \rangle \approx -5$, $\langle \mbh \rangle \approx 10^{7} \msun$, and $\langle \log \mdot \rangle\approx 0$.

\section{Results and Discussion} \label{sec:discussion3}

We now examine our sample in the context of the jet scaling model described above, along with alternative scenarios and notable caveats.

\subsection{No Sign of Domination by Compact Jets}\label{subsec:p3_jets}

\begin{figure}[t]
\epsscale{1.15}
\plotone{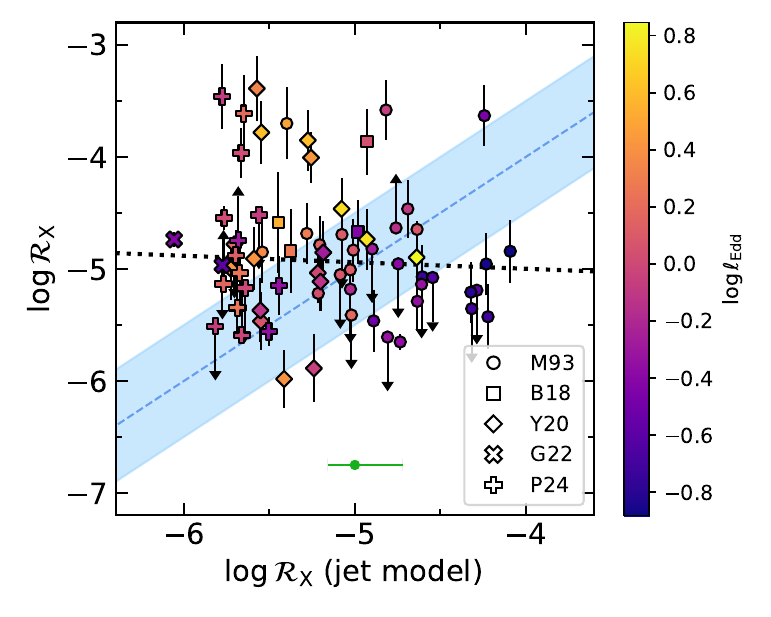}
\caption{$\fweak$-corrected $\log \Rx$ of objects in our sample vs.\ the model $\log \Rx$ calculated as a function of their $\mbh$ and $\mdot$ (Eq.~\ref{eq:FP_LRLX}). The model has been normalized to our sample mean values of $\langle \log \Rx \rangle \approx -5$, $\langle \mbh \rangle \approx 10^{7} \msun$, and $\langle \log \mdot \rangle\approx 0$. The sample is shown by symbols given in the legend and is colorized by $\log \ledd$ (color bar). 1$\sigma$ errors were derived using the Monte Carlo scheme described in Section \ref{subsec:data_err} (arrows denote limits from nondetections). The green dot and bars to the left of the legend illustrate the average uncertainty for modeled $\Rx$. The blue dashed line and shaded area show the 1:1 relationship between the axes with a 1~dex scatter. Were radio emission dominated by a jet, we would expect the sample to follow the 1:1 slope. Instead, the distribution of $\Rx$ in our sample is effectively flat (with large scatter; see Section \ref{sec:discussion3}), as illustrated by the black dotted line.
\label{fig:p3_jetmodels}
}
\end{figure}

Based on the following analysis, it appears that compact jets do not dominate radio emission from our sample of radio-quiet, high/super-Eddington AGNs. Figure \ref{fig:p3_jetmodels} compares $\Rx$ ($\fweak$-corrected per the rule set in Section \ref{subsec:data_lum}) for our sample against the radio jet model ($\Rx$ as a function of our sample's $\mbh$ and $\mdot$ estimates; Eq.~\ref{eq:FP_LRLX} above). The dashed blue line and shaded area show the 1:1 relationship for the axes with a $\pm 0.5$~dex error. Note that while this line crosses our sample due to normalization, our sample crucially does not follow its slope. A Kendall's $\tau$ correlation test gives $p_{\rm null} = 0.5 \pm 0.3$ (i.e., likely no correlation) between the observed and modeled values of $\Rx$, and an ordinary-least-squares fit gives an effectively flat slope ($m = -0.06 \pm 0.10$, where $y=mx+b$). Our sample therefore fails to follow the model we have derived for jet-dominated radio emission, by orders of magnitude, despite our range in both $\Rx$ and $\mbh$. For completeness, we reiterate that altering our choice of bolometric correction or X-ray weakness correction scheme (according to the matrix of 9 correction scenarios discussed in Section \ref{subsec:data_mass_edd}) does not substantially change the above result, nor does adopting a steep slope ($\alphar=-0.7$) for our jet model. 

\subsection{Alternative Sources of Radio Emission} \label{subsec:p3_corona}

We naturally cannot dismiss the possible presence of compact jets on an individual basis, but on a statistical level our sample appears inconsistent with domination by jets. Radio emission from coronal activity presents probably the most simple alternative to examine: $\Rx \sim$\, constant. \citet{Guedel93} found that coronally active stars typically show a ratio $\log \Rx \sim -5$. It has since been widely suggested that radio-quiet AGNs generally follow this relationship, albeit with large scatter \citep[e.g.,][]{Doi15, Jarvis21, Richards21, Yao21, Yang22, Chen23, Ricci23b, Wang23, Shuvo23, Magno25, Njeri25, Shablovinskaia25, Hankla25}. The distribution of $\Rx$ in our sample is consistent with these prior studies.

On the other hand, diagnosis of radio emission from winds or other non-collimated outflows (such as a failed jet) via $\Rx$ poses a significant challenge. While bremsstrahlung free-free emission from a disk wind may not be the dominant contributor of either radio or X-rays \citep{Steenbrugge11}, recent studies indicate that wind shocks may produce synchrotron radio emission on the order $\nu L_{\nu} \sim 10^{-5} \Lbol$, which could become dominant at the $\lesssim$\,1~kpc physical scales covered by our sample's VLA observations \citep[e.g.,][]{Nims15, Yamada24}, particularly at high accretion rates \citep[e.g.,][]{Laor19, Alhosani22, Chen24a, Chen24b, Mestici24}.\footnote{\citet{Chen24a} show a link between disk winds (probed via blueshifted \ion{C}{4} $\lambda$1549) and radio emission in quasars at high $\ledd$, but it remains to be determined whether lower-mass AGNs (such as those in the P24 or G22 subsamples) are capable of efficiently driving disk winds \citep[e.g.,][]{Giustini19, Naddaf22}. Rest-frame UV spectra covering \ion{C}{4} in low-mass samples will be extremely valuable in extending this link down the mass range.} Drawing on the work of, e.g., \citet{Wright75}, \citet{Blustin09}, \citet{King13}, and \citet{Nims15}, it is possible to find that the shocked ambient medium may display $\Rx \propto \mbh^{1/3} \mdot^{(4/3)-q}$. However, there are additional characteristics of the wind and ambient medium that may depend on the black hole mass and/or accretion rate and for which we still lack good constraints. Moreover, we would ideally use the spectral index as part of our radio emission diagnosis, but we lack robust $\alphar$ values for many of the lower-mass and fainter objects.

\begin{figure}[t]
\epsscale{1.15}
\plotone{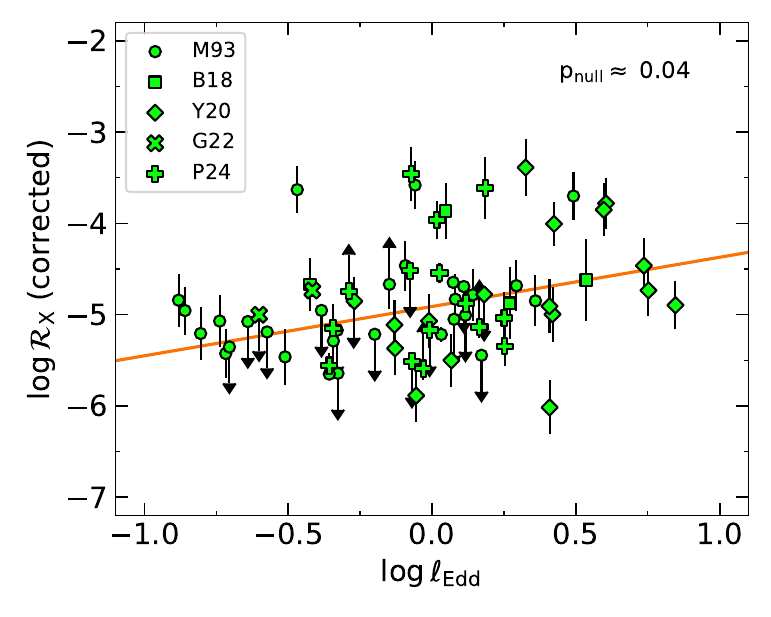}
\caption{Comparison of $\log \Rx$ with $\log \ledd$. The sample is shown by symbols given in the legend. 1$\sigma$ errors were derived using the Monte Carlo scheme described in Section \ref{subsec:data_err} (arrows denote limits from nondetections). The orange line shows an ordinary-least-squares fit to our sample, with a correlation confidence level of $p_{\rm null} = 0.04 \pm 0.03$. We do not expect this fit to extend to lower values of $\ledd$; it is more likely that $\Rx$ is generally flat for $\log \ledd \lesssim -0.5$ but increases (albeit with greater scatter) at higher $\ledd$.
\label{fig:p3_lrlx_ledd}}
\end{figure}

For now, we attempt to explore a link between excess radio emission and outflows at high accretion rates by comparing $\Rx$ and $\ledd$, as shown in Figure \ref{fig:p3_lrlx_ledd}. Using the Monte Carlo scheme described in Section \ref{subsec:data_err}, we perform a Kendall's $\tau$ test and an ordinary-least-squares fit to our sample, finding a correlation statistic $0.20 \pm 0.04$ with $p_{\rm null} = 0.04 \pm 0.03$ and a best-fit slope $m = 0.54 \pm 0.13$. While this result implies no statistically significant correlation, it does at least appear that higher-$\Rx$ objects may be distributed towards higher values of $\ledd$ (albeit with greater scatter in $\Rx$). Importantly, we would not expect this fit to extend to lower values of $\ledd$, as it leads to unexpectedly low $\Rx$ values. Rather than showing a direct correlation, this may simply exemplify the suggestions that radio-quiet AGN cores are dominated initially by coronal activity with a flat $\Rx$ and become more likely to produce excess radio emission from strong outflows as $\ledd$ increases beyond the Eddington limit.\footnote{Variability may contribute to $\Rx$ scatter, but we do not expect it to dominate Figure \ref{fig:p3_jetmodels}, as it would need to preferentially increase (decrease) the observed $\Rx$ at low (high) model $\Rx$.} However, as discussed in Section \ref{subsec:samp_prop3}, our sample may be incomplete, so it is possible that we are missing lower-$\Rx$ objects at high $\ledd$ (or vice versa) due to, e.g., sensitivity limits in existing observations. Deeper samples with more high Eddington-ratio objects will help rule out such a selection bias. Still, we note that a forthcoming study of Swift/BAT-selected AGNs \citep{Baumgartner13} by S. Venselaar et al.\ (in preparation) shows a result similar to ours, using new Atacama Large Millimeter/submillimeter Array (ALMA) observations at 100~GHz along with quasi-simultaneous Swift X-ray observations.

\subsection{Additional Caveats} \label{subsec:p3_caveats}

Perhaps most importantly, our jet model does not account for the possibility that as the inner accretion disk transitions into the expanded and advection-dominated slim phase, the magnetic field and/or the fraction of jet power to disk power \citep[e.g.,][]{Falcke95} may be significantly altered (requiring different values for $\xi_{\rm M}$ and $\xi_{\mdot}$), or that the scale invariance of Eq.~(\ref{eq:FP_LR}) may be somehow broken at higher accretion rates  (although we note that samples including lower-$\ledd$ objects are found to be similarly inconsistent with jet domination; see Section \ref{subsec:FP} below). Furthermore, with the possibility of variable or cyclic behavior around the thin--slim disk state transition \citep[e.g.,][]{Abramowicz88, Devi24}, activity such as outbursts, instability, and state transitions could have an impact on both radio and X-ray production in individual objects. Recent observations have not only shown rapid and extreme radio variability in some objects \citep[e.g.,][]{Jarvela24}, but also pointed to the possibility of episodic magnetic flux accumulation and temporary jet production \citep{Wang25}.

Indeed, higher-resolution studies of radio emission from radio-quiet AGNs (including some objects in our sample) via very long baseline interferometry (VLBI) tend to show a mix of sources with evidence of outflow or jet activity and sources that remain compact/unresolved \citep[which may contribute to the scatter found in $\Rx$; e.g.,][]{Chen23, Wang23}. For example, \citet{Yao21} and \citet{Yang24} provide in-depth VLBI examinations of, respectively, the radio-quiet, near-Eddington NLS1 Mrk 335 and the radio-quiet, super-Eddington PG quasar I~Zw~1 (a.k.a.\ PG 0050+124) and find that these objects show evidence of small (pc-scale) jet activity at 1.5~GHz. Meanwhile, \citet{Yang22} examine four low-mass AGNs (three of which are high-Eddington) and suggest their radio emission comes from a combination of coronal activity and outflows (although we note that extremely compact radio jets could remain unresolved even at VLBI resolutions; e.g., \citealt{Wang23}). By design, existing VLBI studies tend to be biased towards the radio-brighter objects in our sample. Because we still lack sufficient coverage of our sample with VLBI, with particularly critical deficits in the lower-mass and radio-fainter regimes, we caution against extrapolating the existing observations to the whole population. Combined-array (e.g., High Sensitivity Array) radio observations can help overcome this limitation.

\begin{figure*}[t]
\gridline{\fig{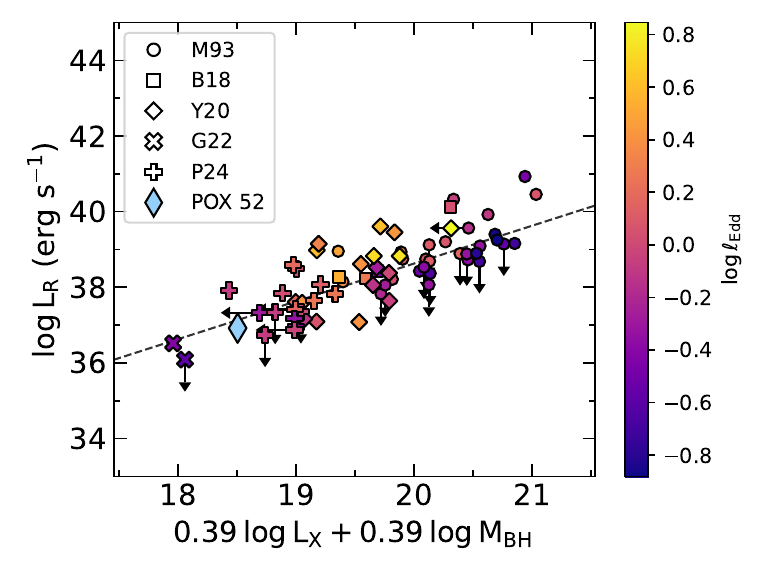}{0.49\textwidth}{\textbf{(a)}}
        \fig{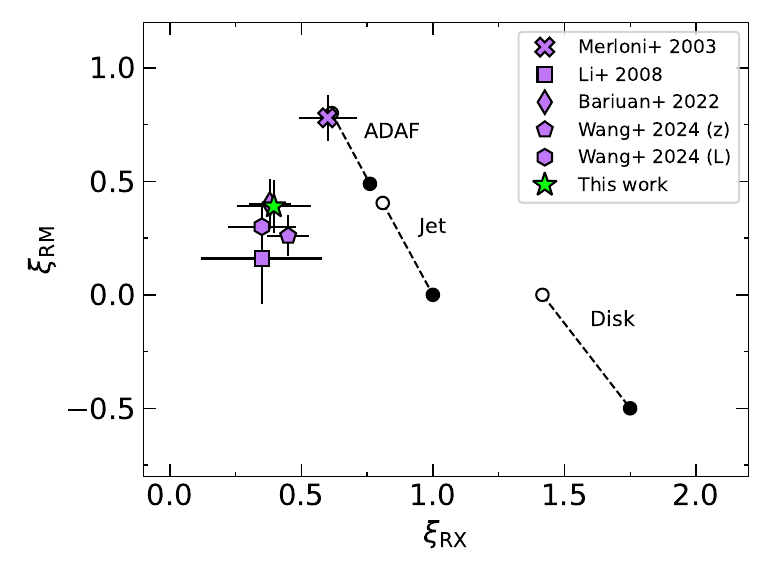}{0.49\textwidth}{\textbf{(b)}}}
\caption{Our sample examined in the context of the fundamental plane of black hole accretion. (a): Result of an ordinary-least-squares fit (dashed line) to our sample, using our Monte Carlo scheme to account for error and nondetections (Section \ref{subsec:data_err}). Our sample is shown by symbols given in the legend (categorized by subsample; Section \ref{subsec:samp3}) and is colorized by $\ledd$ (color bar). The light blue diamond shows POX 52, a massive black hole ($\mbh \sim 1.6 \times 10^{5} \msun$) which was recently detected in radio for the first time by \citet{Yuan25}. Arrows denote limits from radio or X-ray nondetections.
(b): Fit parameters $\xi_{\rm RX}$ and $\xi_{\rm RM}$ for our sample (green star) compared to a number of other works in the literature (purple symbols), including \citet{Merloni03}; radio-quiet objects at $z < 0.13$ from \citet{Li08}; radio-quiet objects at $z < 1.5$ from \citet{Bariuan22}; and radio-quiet objects from \citet{Wang24}, limited by $0.1 < z < 1$ (denoted ``Wang+ 2024 (z)'', including both low- and high-$\ledd$ objects) or by $\ledd > 0.03$ (denoted ``Wang+ 2024 (L)'', including their full range of $0.1 < z < 4$). Model-predicted ranges of the fit parameters are shown by circles with connecting dashed lines that trace variation in the radio spectral index (open circles assume a flat $\alphar = 0$ and filled circles assume a steep $\alphar = -0.5$) for three origins of X-ray emission: advection-dominated accretion flow (ADAF), jet, and standard disk/corona (e.g., Sections 5.2 and 5.3 of \citealt{Merloni03}). Note that all three scenarios interpret radio emission as originating from a synchrotron jet, and all of the radio-quiet samples are inconsistent with these scenarios.
\label{fig:p3_FP}}
\end{figure*}

Finally, we have treated star formation as an insignificant contributor of radio emission. At 5--10~GHz, we do not expect star formation activity in our sample to dominate over AGN activity at $\lesssim$\,1~kpc, particularly for the more massive and brighter nuclei. The lower-mass objects in the P24 subsample have been found to be consistent with domination by AGN activity \citep{Paul24}. There remains the issue of possible nuclear star formation (e.g., see Section 4.4 of \citealt{Marziani25}), but this is still not well understood. 

\subsection{Comparison to the Fundamental Plane of Black Hole Activity} \label{subsec:FP}

While the fundamental plane was developed for low Eddington-ratio black holes (including hard-state black hole XRBs; see Section \ref{sec:intro3}), for completeness we briefly assess our sample in its context. We apply the fit:
\begin{equation} \label{eq:FP_fit}
\log \Lr\ =\ \xi_{\rm RX} \log \Lx\ +\ \xi_{\rm RM} \log \mbh\ +\ b_{\rm r},
\end{equation}
where $\xi_{\rm RX}$ and $\xi_{\rm RM}$ are as defined by Eq.~(11) of \citet{Merloni03} and $b_{\rm r}$ is a normalization constant. Using the Monte Carlo scheme described in Section \ref{subsec:data_err}, we find $\xi_{\rm RX} = 0.39 \pm 0.14$, $\xi_{\rm RM} = 0.39 \pm 0.12$, and $b_{\rm r} = 18.63 \pm 5.29$. We find that this fit is consistent (within the range of error) for all choices of X-ray weakness correction or bolometric correction scheme (Sections \ref{subsec:data_lum}, \ref{subsec:data_mass_edd}). Note that we do not include black hole XRBs because there no known analogs with $\ledd \geq 0.1$ and compact core radio emission.\footnote{While there may be a small number of coordinated radio and X-ray observations of XRB outbursts in comparable high-$\ledd$ states, identifying XRB states that are convincingly analogous to our sample is outside the scope of this paper  (see the 2015 outburst of V404 Cygni as a possible example; \citealt{Motta17}).}

Figure \ref{fig:p3_FP}(a) shows the result of our fit, and Figure \ref{fig:p3_FP}(b) compares the fit parameters $\xi_{\rm RX}$ and $\xi_{\rm RM}$ to a number of prior studies involving radio-quiet AGNs on the fundamental plane. Using Eq.~(11) of \citet{Merloni03} along with our fit parameters to solve for the X-ray scaling power $q$ and the jet electron distribution power-law index $p$ (see Section \ref{sec:models}), we find values of $q \approx 1$ and $p \approx -16$. This value of $q$ is reasonable for efficient X-ray emission dominated by the accretion flow and corona (though it is somewhat large for the radiation pressure-dominated scenario; e.g., \citealt{Beckert02, Merloni02, Merloni03}; see also discussion in Appendix \ref{sec:app}). However, the value of $p$ that we find is not physical --- the expectation is $p \sim$\,2--3 for a shock-accelerated plasma \citep[e.g.,][]{Jones91}. Even with $q$ forced to 0.5 to match our model assumptions, we still find $p \approx -8$ from the fundamental plane fit. Performing the same for the other radio-quiet samples shown in Figure \ref{fig:p3_FP}(b) (with similar values of $\xi_{\rm RM}$ and $\xi_{\rm RX}$) will likely produce similarly non-physical values. We infer from the above that radio-quiet, high-Eddington AGNs are not consistent with the interpretation of jet-dominated radio emission that is usually invoked for the fundamental plane, and a high level of caution is advisable when attempting to place high-$\ledd$ massive black hole candidates on the fundamental plane for the purpose of mass estimation \citep[e.g.,][]{Panurach24}.

\section{Summary and Conclusions} \label{sec:summ3}

We have gathered a moderately sized sample ($N = 69$) of radio-quiet, high Eddington-ratio ($\ledd \geq 0.1$) AGNs across a wide range of masses ($\mbh \sim 10^{5}$--$10^{9} \msun$, notably extending down towards the massive black hole regime). We compared the ratios of their 5~GHz radio to 2--10~keV X-ray luminosities ($\Rx = \lrlx$) to a model where both the radio emission of a compact jet and accretion-powered X-ray emission scale with black hole mass and accretion rate. Many objects in this sample appear to be accreting at super-Eddington rates and may therefore be in a slim-disk accretion state, wherein a radiatively inefficient, advection-dominated inner disk may produce saturated optical--UV luminosities and the shielding of X-rays from our line of sight (Section \ref{sec:intro3}). Accordingly, we modified our adopted formalisms to account for these effects (Section \ref{sec:models}). We summarize our findings as follows:

\begin{itemize}
\setlength{\itemsep}{0pt}
    \item  The observed $\Rx$ across this sample is inconsistent with the radio jet model, suggesting that on average, radio emission in radio-quiet, high-Eddington AGNs is not dominated by a compact jet (Section \ref{subsec:p3_jets}). Individual objects may still display weak or intermittent jet activity (Section \ref{subsec:p3_caveats}). 
    \item This sample is generally consistent with prior observations indicating radio-quiet AGNs have a roughly constant ratio $\log \Rx \sim -5$, similar to the \citet{Guedel93} relationship for coronally active stars (Section \ref{subsec:p3_corona}).
    \item Recent observations suggest that strong, non-collimated outflows such as winds are commonly driven at high $\ledd$, and these outflows likely contribute excess radio emission. We may then expect to see a flat $\Rx$ at lower $\ledd$ that becomes a positive correlation at higher $\ledd$. Qualitative inspection of our sample suggests this may be the case, though we stress that our statistical analysis is highly inconclusive (Section \ref{subsec:p3_corona}). Larger samples are needed to more completely probe the high-Eddington population of radio-quiet AGNs and determine whether the apparent lack of lower-$\Rx$ objects at high $\ledd$ is real.
\end{itemize}

To address the above points, continued studies of core radio and X-ray activity (e.g., higher-resolution radio observations and more robust spectral index estimates, polarimetry to help distinguish emission mechanisms/geometries, etc.) and optical--UV outflow signatures (e.g., coverage of \ion{C}{4} to examine winds in low-mass AGNs) will help us unravel the outflow properties of radio-quiet AGNs. Upcoming planned/proposed observatories such as a next-generation VLA \citep[ngVLA;][]{Murphy18} or Advanced X-ray Imaging Satellite \citep[AXIS;][]{AXIS23} will greatly improve our capabilities.

\begin{acknowledgments}
We thank the anonymous referee for constructive comments that helped improve this manuscript. We thank Emilia J{\"a}rvel{\"a} for valuable discussion and suggestions. We acknowledge support from the National Science Foundation under grant number 2206123. This research has made use of data obtained from the Chandra Source Catalog, provided by the Chandra X-ray Center (CXC). It has also made use of data obtained from eROSITA, the soft X-ray instrument aboard SRG, a joint Russian-German science mission supported by the Russian Space Agency (Roskosmos), in the interests of the Russian Academy of Sciences represented by its Space Research Institute (IKI), and the Deutsches Zentrum für Luft- und Raumfahrt (DLR). The SRG spacecraft was built by Lavochkin Association (NPOL) and its subcontractors, and is operated by NPOL with support from the Max Planck Institute for Extraterrestrial Physics (MPE). The development and construction of the eROSITA X-ray instrument was led by MPE, with contributions from the Dr.\ Karl Remeis Observatory Bamberg \& ECAP (FAU Erlangen-Nuernberg), the University of Hamburg Observatory, the Leibniz Institute for Astrophysics Potsdam (AIP), and the Institute for Astronomy and Astrophysics of the University of Tübingen, with the support of DLR and the Max Planck Society. The Argelander Institute for Astronomy of the University of Bonn and the Ludwig Maximilians Universität Munich also participated in the science preparation for eROSITA. This research also made use of the following resources: the SIMBAD database and the VizieR catalogue access tool, both operated at CDS, Strasbourg, France, and the NASA/IPAC Extragalactic Database (NED), which is funded by the National Aeronautics and Space Administration and operated by the California Institute of Technology.
\end{acknowledgments}

\facilities{CDS, CXO, eROSITA, NED, Swift, VLA, XMM}

\software{Astropy \citep{astropy13, astropy18}, NumPy \citep{Numpy}, SciPy \citep{scipy}, pandas \citep{pandas}, Pingouin \citep{Pingouin}, TOPCAT \citep{topcat}, Ned Wright's JavaScript cosmology calculator \citep{Wright06}}

\pagebreak

\appendix
\restartappendixnumbering

\section{Data Tables}\label{sec:app_tables}

\startlongtable
\begin{deluxetable*}{lccccccc}
\tablenum{A1}
\tablecaption{Sample Optical Properties\label{tab:P3sampOPT}}
\tablehead{
\colhead{Source Name}  &  \colhead{Subsamp.}  &  \colhead{$z$}  &  \colhead{$\log L_{\rm 5100}$}  &  \colhead{FWHM}  &  \colhead{$\log \mbh$}  & \colhead{$\ledd$}  &  \colhead{Ref.}  \\
\colhead{}  &  \colhead{}  &  \colhead{}  &  \colhead{(erg s$^{-1}$)}  &  \colhead{(km s$^{-1}$)}  &  \colhead{($\msun$)}  &  \colhead{}  &  \colhead{}  \\
\colhead{(1)}  &  \colhead{(2)}  &  \colhead{(3)}  &  \colhead{(4)}  &  \colhead{(5)}  &  \colhead{(6)}  &  \colhead{(7)}  &  \colhead{(8)} 
}
\startdata
PG 0026+129 & M93 & 0.142 & 44.97 & 1860$^{\beta}$ & 7.71 & 1.49 & D11 \\
PG 0050+124 & M93 & 0.061 & 44.39 & 1240$^{\beta}$ & 7.12 & 1.96 & D11 \\
PG 0052+251 & M93 & 0.155 & 44.98 & 5200$^{\beta}$ & 8.60 & 0.19 & D11 \\
PG 0157+001 & M93 & 0.164 & 45.00 & 2460$^{\beta}$ & 7.96 & 0.87 & D11 \\
PG 0804+761 & M93 & 0.100 & 44.77 & 3070$^{\beta}$ & 8.06 & 0.45 & D11 \\
PG 0844+349 & M93 & 0.064 & 44.29 & 2420$^{\beta}$ & 7.66 & 0.47 & D11 \\
PG 0921+525 & M93 & 0.035 & 43.54 & 2120$^{\beta}$ & 7.24 & 0.31 & D11 \\
PG 0947+396 & M93 & 0.206 & 45.18 & 4830$^{\beta}$ & 8.62 & 0.27 & D11 \\
PG 1001+054 & M93 & 0.161 & 44.67 & 1740$^{\beta}$ & 7.53 & 1.29 & D11 \\
PG 1011$-$040 & M93 & 0.058 & 44.06 & 1440$^{\beta}$ & 7.12 & 1.08 & D11 \\
PG 1012+008 & M93 & 0.185 & 44.93 & 2640$^{\beta}$ & 7.99 & 0.71 & D11 \\
PG 1114+445 & M93 & 0.144 & 44.73 & 4570$^{\beta}$ & 8.39 & 0.20 & D11 \\
PG 1115+407 & M93 & 0.154 & 44.56 & 1720$^{\beta}$ & 7.47 & 1.19 & D11 \\
PG 1116+215 & M93 & 0.177 & 45.29 & 2920$^{\beta}$ & 8.23 & 0.81 & D11 \\
PG 1119+120 & M93 & 0.049 & 43.99 & 1820$^{\beta}$ & 7.29 & 0.63 & D11 \\
PG 1126$-$041 & M93 & 0.060 & 44.17 & 2150$^{\beta}$ & 7.51 & 0.53 & D11 \\
PG 1149$-$110 & M93 & 0.049 & 43.77 & 3060$^{\beta}$ & 7.66 & 0.18 & D11 \\
PG 1202+281 & M93 & 0.165 & 44.56 & 5050$^{\beta}$ & 8.41 & 0.14 & D11 \\
PG 1211+143 & M93 & 0.085 & 44.83 & 1860$^{\beta}$ & 7.65 & 1.31 & D11 \\
PG 1216+069 & M93 & 0.334 & 45.60 & 5190$^{\beta}$ & 8.85 & 0.34 & D11 \\
PG 1229+204 & M93 & 0.064 & 44.22 & 3360$^{\beta}$ & 7.92 & 0.23 & D11 \\
PG 1244+026 & M93 & 0.048 & 43.68 & 830$^{\beta}$ & 6.49 & 2.29 & D11 \\
PG 1402+261 & M93 & 0.164 & 44.80 & 1910$^{\beta}$ & 7.66 & 1.21 & D11 \\
PG 1404+226 & M93 & 0.098 & 44.14 & 880$^{\beta}$ & 6.72 & 3.10 & D11 \\
PG 1411+442 & M93 & 0.089 & 44.43 & 2670$^{\beta}$ & 7.80 & 0.44 & D11 \\
PG 1415+451 & M93 & 0.114 & 44.32 & 2620$^{\beta}$ & 7.74 & 0.41 & D11 \\
PG 1416$-$129 & M93 & 0.129 & 44.92 & 6110$^{\beta}$ & 8.72 & 0.13 & D11 \\
PG 1435$-$067 & M93 & 0.129 & 44.88 & 3180$^{\beta}$ & 8.13 & 0.47 & D11 \\
PG 1440+356 & M93 & 0.077 & 44.35 & 1450$^{\beta}$ & 7.24 & 1.39 & D11 \\
PG 1626+554 & M93 & 0.133 & 44.44 & 4490$^{\beta}$ & 8.26 & 0.16 & D11 \\
PG 2112+059 & M93 & 0.466 & 45.90 & 3190$^{\beta}$ & 8.55 & 1.18 & D11 \\
FBQS J0752+2617 & B18 & 0.082 & 43.66 & 1337$^{\beta}$ & 6.89 & 1.86 & Ra17 \\
IRAS 13349+2438 & B18 & 0.108 & 44.36 & 2800$^{\beta}$ & 7.82 & 1.12 & G06 \\
Mrk 705 & B18 & 0.029 & 43.20 & 2165$^{\beta}$ & 7.13 & 0.38 & Ra17 \\
RX J0957.1+2433 & B18 & 0.082 & 43.57 & 925$^{\beta}$ & 6.54 & 3.43 & Ra17 \\
Ark 564 & Y20 & 0.025 & 43.25 & 860$^{\beta}$ & 6.34 & 2.56 & W13 \\
IRAS 04416+1215 & Y20 & 0.089 & 44.06 & 1470$^{\beta}$ & 7.13 & 2.66 & W13 \\
IRASF 12397+3333 & Y20 & 0.043 & 43.05 & 1640$^{\beta}$ & 6.83 & 0.54 & W13 \\
KUG 1031+398 & Y20 & 0.042 & 43.24 & 940$^{\beta}$ & 6.42 & 2.11 & W13 \\
Mrk 1044 & Y20 & 0.016 & 42.91 & 1010$^{\beta}$ & 6.35 & 1.17 & W13 \\
Mrk 142 & Y20 & 0.045 & 43.40 & 1620$^{\beta}$ & 6.95 & 0.88 & W13 \\
Mrk 335 & Y20 & 0.026 & 43.55 & 1710$^{\beta}$ & 7.06 & 0.98 & W13 \\
Mrk 42 & Y20 & 0.025 & 42.48 & 940$^{\beta}$ & 6.11 & 0.74 & W13 \\
Mrk 478 & Y20 & 0.077 & 44.39 & 1270$^{\beta}$ & 7.14 & 5.65 & W13 \\
Mrk 486 & Y20 & 0.039 & 43.32 & 1680$^{\beta}$ & 6.95 & 0.74 & W13 \\
Mrk 684 & Y20 & 0.046 & 43.67 & 1150$^{\beta}$ & 6.77 & 2.56 & W13 \\
Mrk 957 & Y20 & 0.071 & 43.26 & 690$^{\beta}$ & 6.16 & 4.02 & W13 \\
Nab 0205+024 & Y20 & 0.156 & 44.70 & 1410$^{\beta}$ & 7.36 & 7.02 & W13 \\
PG 1448+273 & Y20 & 0.065 & 44.09 & 1050$^{\beta}$ & 6.85 & 5.45 & W13 \\
RX J1355.2+5612 & Y20 & 0.122 & 43.92 & 1100$^{\beta}$ & 6.83 & 3.95 & W13 \\
SDSS J010712.04+140845.0 & Y20 & 0.076 & 42.75 & 790$^{\beta}$ & 6.07 & 1.52 & W13 \\
SDSS J114008.71+030711.4 & Y20 & 0.081 & 42.93 & 680$^{\beta}$ & 6.01 & 2.62 & W13 \\
SDSS J085125.81+393541.7 & G22 & 0.041 & 40.91 & 894$^{\alpha}$ & 5.01 & 0.25 & R13 \\
SDSS J095418.15+471725.1 & G22 & 0.033 & 40.64 & 636$^{\alpha}$ & 4.56 & 0.38 & R13 \\
SDSS J032515.58+003408.4 & P24 & 0.102 & 42.54 & 970$^{\alpha}$ & 6.05 & 0.98 & S11 \\
SDSS J091449.05+085321.1 & P24 & 0.140 & 42.89 & 849$^{\alpha}$ & 6.14 & 1.79 & S11 \\
SDSS J092438.88+560746.8 & P24 & 0.025 & 42.10 & 1180$^{\alpha}$ & 5.96 & 0.44 & S11 \\
SDSS J094310.11+604559.1 & P24 & 0.074 & 42.57 & 807$^{\alpha}$ & 5.90 & 1.46 & S11 \\
SDSS J101246.59+061604.7 & P24 & 0.078 & 42.53 & 941$^{\alpha}$ & 6.02 & 1.04 & S11 \\
SDSS J105755.66+482501.9 & P24 & 0.073 & 41.82 & 957$^{\alpha}$ & 5.61 & 0.51 & S11 \\
SDSS J124035.82$-$002919.4 & P24 & 0.081 & 42.88 & 915$^{\alpha}$ & 6.20 & 1.53 & S11 \\
SDSS J131651.29+055646.9 & P24 & 0.055 & 42.27 & 1260$^{\alpha}$ & 6.11 & 0.45 & S11 \\
SDSS J131926.52+105610.9 & P24 & 0.064 & 42.31 & 840$^{\alpha}$ & 5.79 & 1.06 & S11 \\
SDSS J140829.27+562823.4 & P24 & 0.134 & 42.66 & 889$^{\alpha}$ & 6.04 & 1.31 & S11 \\
SDSS J144052.60$-$023506.2 & P24 & 0.044 & 42.44 & 950$^{\alpha}$ & 5.97 & 0.93 & S11 \\
SDSS J155909.62+350147.5 & P24 & 0.031 & 42.92 & 862$^{\alpha}$ & 6.17 & 1.78 & S11 \\
SDSS J162636.40+350242.1 & P24 & 0.034 & 41.98 & 802$^{\alpha}$ & 5.55 & 0.85 & S11 \\
SDSS J163159.59+243740.2 & P24 & 0.043 & 42.07 & 839$^{\alpha}$ & 5.64 & 0.85 & S11 \\
SDSS J165636.98+371439.5 & P24 & 0.063 & 42.58 & 1070$^{\alpha}$ & 6.16 & 0.84 & S11 \\
\enddata
\tablecomments{
Column (1): object name. Column (2): subsample identifier (see Section \ref{subsec:samp3}). Column (3): redshift. Column (4): logarithm of the continuum luminosity $L_{\rm 5100} = \nu L_{\nu}$ at 5100~{\AA} (see Section \ref{subsec:data_lum}). Column (5): FWHM of broad H$\alpha$ (denoted by superscript $\alpha$) or H$\beta$ (superscript $\beta$) emission line. Column (6): logarithm of black hole mass estimate (see Section \ref{subsec:data_mass_edd}). Column (7): Eddington luminosity ratio estimate (see Section \ref{subsec:data_mass_edd}). Column (8): optical data reference. References: D11: \citet{Davis11}; G06: \citet{Gallo06}; R13: \citet{Reines13}; Ra17: \citet{Rakshit17}; S11: \citet{Shen11}; W13: \citet{Wang13}.
}
\end{deluxetable*}

\startlongtable
\begin{deluxetable*}{lccccccccc}
\tablenum{A2}
\tablecaption{Sample Radio and X-ray Properties \label{tab:P3sampRX}}
\tablehead{
\colhead{Source Name}  &  \colhead{Subsamp.}  &  \colhead{$R_{\rm O}$}  &  \colhead{Ratio$_{\rm p/i}$}  &  \colhead{$\log \Lr$}  &  \colhead{$\log \Lx$}  & \colhead{$\fweak$}  &  \colhead{$\log \Rx$}  &  \colhead{Src.}  &  \colhead{Ref.}  \\
\colhead{}  &  \colhead{}  &  \colhead{}  &  \colhead{}  &  \colhead{(erg s$^{-1}$)}  &  \colhead{(erg s$^{-1}$)}  &  \colhead{}  &  \colhead{}  &  \colhead{}  &  \colhead{} \\
\colhead{(1)}  &  \colhead{(2)}  &  \colhead{(3)}  &  \colhead{(4)}  &  \colhead{(5)}  &  \colhead{(6)}  &  \colhead{(7)}  &  \colhead{(8)}  &  \colhead{(9)}  &  \colhead{(10)}  
}
\startdata
PG 0026+129 & M93 & $<$\,0.10 & \nodata & $<$\,38.89 & 44.34 & 0.47 & $<$\,$-$5.45 & S & Ri17 \\
PG 0050+124 & M93 & 0.42 & 0.8$^{*}$ & 38.93 & 43.61 & 1.59 & $-$4.68$\pm$0.28 & X & B09 \\
PG 0052+251 & M93 & 0.17 & 0.6$^{*}$ & 39.16 & 44.59 & 0.43 & $-$5.43$\pm$0.29 & X & B09 \\
PG 0157+001 & M93 & 2.39 & 0.8$^{*}$ & 40.33 & 43.91 & 1.20 & $-$3.58$\pm$0.27 & C & CSC \\
PG 0804+761 & M93 & 0.25 & 0.4$^{*}$ & 39.09 & 44.38 & 0.39 & $-$5.29$\pm$0.27 & X & B09 \\
PG 0844+349 & M93 & $<$\,0.07 & \nodata & $<$\,38.07 & 43.71 & 0.73 & $<$\,$-$5.64 & C & S10 \\
PG 0921+525 & M93 & 0.94 & 0.5$^{*}$ & 38.43 & 43.89 & 0.25 & $-$5.46$\pm$0.27 & X & B09 \\
PG 0947+396 & M93 & $<$\,0.10 & \nodata & $<$\,39.14 & 44.33 & 0.67 & $<$\,$-$5.19 & C & CSC \\
PG 1001+054 & M93 & $<$\,0.32 & \nodata & $<$\,39.12 & 41.88 & 86.78 & $<$\,$-$4.69 & C & CSC \\
PG 1011$-$040 & M93 & 0.17 & 1.5$^{*}$ & 38.21 & 42.41 & 10.41 & $-$5.22$\pm$0.09 & E & ER \\
PG 1012+008 & M93 & 0.48 & 0.8$^{*}$ & 39.56 & $<$\,44.23 & $>$\,0.56 & $>$\,$-$4.67 & E & ER \\
PG 1114+445 & M93 & $<$\,0.10 & \nodata & $<$\,38.68 & 44.04 & 1.00 & $<$\,$-$5.36 & X & B09 \\
PG 1115+407 & M93 & $<$\,0.17 & \nodata & $<$\,38.74 & 43.80 & 0.85 & $<$\,$-$5.05 & X & B09 \\
PG 1116+215 & M93 & 0.48 & 0.7$^{*}$ & 39.92 & 44.39 & 1.05 & $-$4.46$\pm$0.25 & X & B09 \\
PG 1119+120 & M93 & $<$\,0.08 & \nodata & $<$\,37.83 & 43.04 & 2.19 & $<$\,$-$5.22 & E & ER \\
PG 1126$-$041 & M93 & $<$\,0.09 & \nodata & $<$\,38.06 & 42.92 & 3.82 & $<$\,$-$4.86 & E & ER \\
PG 1149$-$110 & M93 & 0.69 & 0.5$^{*}$ & 38.53 & 43.60 & 0.48 & $-$5.07$\pm$0.30 & S & Ri17 \\
PG 1202+281 & M93 & 0.78 & 0.8$^{*}$ & 39.4 & 44.36 & 0.38 & $-$4.96$\pm$0.26 & X & B09 \\
PG 1211+143 & M93 & $<$\,0.09 & \nodata & $<$\,38.69 & 43.70 & 2.27 & $<$\,$-$5.01 & X & B09 \\
PG 1216+069 & M93 & 2.19 & 1.3$^{*}$ & 40.93 & 44.56 & 1.69 & $-$3.63$\pm$0.24 & X & B09 \\
PG 1229+204 & M93 & $<$\,0.17 & \nodata & $<$\,38.37 & 43.44 & 1.50 & $<$\,$-$5.08 & X & B09 \\
PG 1244+026 & M93 & 0.35 & 0.6$^{*}$ & 38.14 & 42.99 & 1.32 & $-$4.85$\pm$0.30 & X & B09 \\
PG 1402+261 & M93 & 0.29 & 0.7$^{*}$ & 39.2 & 44.03 & 1.28 & $-$4.83$\pm$0.27 & X & B09 \\
PG 1404+226 & M93 & 0.77 & 0.8$^{*}$ & 38.95 & 42.65 & 1.82 & $-$3.7$\pm$0.28 & C & CSC \\
PG 1411+442 & M93 & 0.23 & 0.9$^{*}$ & 38.72 & 42.92 & 28.45 & $-$5.65$\pm$0.07 & X & P05 \\
PG 1415+451 & M93 & $<$\,0.19 & \nodata & $<$\,38.53 & 43.49 & 1.65 & $<$\,$-$4.95 & X & B09 \\
PG 1416$-$129 & M93 & 0.25 & 0.2$^{*}$ & 39.25 & 44.09 & 1.43 & $-$4.84$\pm$0.30 & X & B09 \\
PG 1435$-$067 & M93 & $<$\,0.12 & \nodata & $<$\,38.88 & 44.05 & 0.79 & $<$\,$-$5.17 & E & ER \\
PG 1440+356 & M93 & 0.30 & 0.5$^{*}$ & 38.74 & 43.53 & 1.36 & $-$4.78$\pm$0.25 & X & B09 \\
PG 1626+554 & M93 & 0.33 & 2.1$^{*}$ & 38.9 & 44.10 & 0.41 & $-$5.21$\pm$0.30 & X & P05 \\
PG 2112+059 & M93 & 0.35 & 0.9$^{*}$ & 40.46 & 44.07 & 10.70 & $-$4.65$\pm$0.08 & C & CSC \\
FBQS J0752+2617 & B18 & 0.44 & 1.0 & 38.23 & 43.11 & 1.17 & $-$4.88$\pm$0.40 & E & ER \\
IRAS 13349+2438 & B18 & 7.23 & 0.9 & 40.12 & 43.99 & 0.50 & $-$3.86$\pm$0.32 & C & H07 \\
Mrk 705 & B18 & 2.09 & 0.9 & 38.43 & 43.10 & 0.75 & $-$4.67$\pm$0.27 & C & CSC \\
RX J0957.1+2433 & B18 & 0.60 & 1.0 & 38.27 & 42.89 & 1.68 & $-$4.62$\pm$0.42 & E & ER \\
Ark 564 & Y20 & 2.85 & 1.0 & 38.61 & 43.52 & 0.19 & $-$4.91$\pm$0.26 & X & W13 \\
IRAS 04416+1215 & Y20 & 2.95 & 0.8 & 39.45 & 43.46 & 0.68 & $-$4.01$\pm$0.23 & S & W13 \\
IRASF 12397+3333 & Y20 & 3.60 & 1.1 & 38.52 & 43.38 & 0.25 & $-$4.85$\pm$0.30 & X & W13 \\
KUG 1031+398 & Y20 & 9.84 & 1.0 & 39.15 & 42.54 & 1.93 & $-$3.39$\pm$0.29 & X & W13 \\
Mrk 1044 & Y20 & 0.19 & 1.0 & 37.09 & 42.59 & 1.28 & $-$5.50$\pm$0.30 & X & W13 \\
Mrk 142 & Y20 & 0.21 & 0.9 & 37.64 & 43.53 & 0.29 & $-$5.89$\pm$0.30 & S & W13 \\
Mrk 335 & Y20 & 0.83 & 1.0 & 38.38 & 43.45 & 0.45 & $-$5.07$\pm$0.26 & X & W13 \\
Mrk 42 & Y20 & 0.59 & 1.0 & 37.16 & 42.53 & 0.75 & $-$5.37$\pm$0.30 & S & W13 \\
Mrk 478 & Y20 & 0.32 & 0.5 & 38.83 & 43.56 & 1.04 & $-$4.73$\pm$0.30 & X & W13 \\
Mrk 486 & Y20 & 0.67 & 1.0 & 38.06 & 43.17 & 0.61 & $-$5.11$\pm$0.27 & X & W13 \\
Mrk 684 & Y20 & 0.03 & 1.0 & 37.08 & 43.10 & 1.22 & $-$6.02$\pm$0.28 & X & W13 \\
Mrk 957 & Y20 & 6.25 & 0.8 & 38.98 & 42.76 & 1.17 & $-$3.78$\pm$0.27 & X & W13 \\
Nab 0205+024 & Y20 & 0.82 & 1.0 & 39.56 & 44.46 & 0.21 & $-$4.90$\pm$0.26 & S & W13 \\
PG 1448+273 & Y20 & 0.66 & 1.0 & 38.83 & 43.29 & 1.25 & $-$4.46$\pm$0.27 & X & W13 \\
RX J1355.2+5612 & Y20 & 5.59 & 1.0 & 39.61 & 43.46 & 0.74 & $-$3.85$\pm$0.26 & S & W13 \\
SDSS J010712.04+140845.0 & Y20 & $<$\,0.87 & \nodata & $<$\,37.61 & 42.39 & 1.51 & $<$\,$-$4.78 & X & W13 \\
SDSS J114008.71+030711.4 & Y20 & 0.54 & 0.8 & 37.59 & 42.59 & 1.29 & $-$5.00$\pm$0.32 & X & W13 \\
SDSS J085125.81+393541.7 & G22 & $<$\,1.83 & \nodata & $<$\,36.09 & 41.09 & 2.18 & $<$\,$-$5.00 & C & B17 \\
SDSS J095418.15+471725.1 & G22 & 9.30 & 0.8 & 36.52 & 40.08 & 15.03 & $-$4.74$\pm$0.07 & C & B17 \\
SDSS J032515.58+003408.4 & P24 & $<$\,0.77 & \nodata & $<$\,37.36 & 41.30 & 16.85 & $<$\,$-$5.17 & C & D09 \\
SDSS J091449.05+085321.1 & P24 & 0.98 & 0.6 & 37.82 & 43.17 & 0.36 & $-$5.35$\pm$0.24 & C & G14 \\
SDSS J092438.88+560746.8 & P24 & 0.91 & 0.9 & 36.95 & 41.50 & 10.11 & $-$5.56$\pm$0.13 & C & D12 \\
SDSS J094310.11+604559.1 & P24 & 0.83 & 0.9 & 37.41 & $<$\,40.97 & $>$\,38.06 & $>$\,$-$5.13 & C & G07 \\
SDSS J101246.59+061604.7 & P24 & 10.83 & 0.7 & 38.49 & 42.46 & 1.20 & $-$3.97$\pm$0.23 & C & G14 \\
SDSS J105755.66+482501.9 & P24 & 3.84 & 0.7 & 37.33 & $<$\,41.22 & $>$\,7.08 & $>$\,$-$4.74 & C & D12 \\
SDSS J124035.82$-$002919.4 & P24 & 6.01 & 0.9 & 38.59 & 42.20 & 5.74 & $-$3.61$\pm$0.33 & C & G07 \\
SDSS J131651.29+055646.9 & P24 & 0.96 & 0.4 & 37.17 & 42.32 & 1.73 & $-$5.15$\pm$0.27 & C & D12 \\
SDSS J131926.52+105610.9 & P24 & 4.04 & 1.1 & 37.84 & 41.16 & 16.79 & $-$4.54$\pm$0.11 & C & D12 \\
SDSS J140829.27+562823.4 & P24 & 2.98 & 0.8 & 38.07 & 42.95 & 0.44 & $-$4.88$\pm$0.22 & C & G14 \\
SDSS J144052.60$-$023506.2 & P24 & 0.33 & 0.3 & 36.87 & $<$\,40.84 & $>$\,42.07 & $>$\,$-$5.59 & C & D12 \\
SDSS J155909.62+350147.5 & P24 & 0.65 & 0.8 & 37.64 & 42.68 & 0.93 & $-$5.04$\pm$0.32 & C & D12 \\
SDSS J162636.40+350242.1 & P24 & $<$\,0.70 & \nodata & $<$\,36.73 & 41.42 & 6.76 & $<$\,$-$5.51 & C & D12 \\
SDSS J163159.59+243740.2 & P24 & 8.50 & 1.0 & 37.91 & 41.38 & 5.65 & $-$3.46$\pm$0.31 & C & D12 \\
SDSS J165636.98+371439.5 & P24 & $<$\,0.70 & \nodata & $<$\,37.34 & 41.86 & 5.41 & $<$\,$-$4.52 & C & D12 \\
\enddata
\tablecomments{
Column (1): object name. Column (2): subsample identifier (see Section \ref{subsec:samp3}). Column (3): radio-to-optical flux density ratio $\Ro=f\textsubscript{5 GHz}/f\textsubscript{4400 {\AA}}$, where $f\textsubscript{\rm 5 GHz}$ and $f\textsubscript{4400 {\AA}}$ are, respectively, the radio and optical flux densities at 5~GHz and 4400~{\AA} \citep[e.g.,][]{Kellermann89,Stocke92}. Column (4): ratio of peak to integrated VLA radio flux densities (see Section \ref{subsec:samp3}). Column (5): logarithm of radio luminosity $L_{\rm R} = \nu L_{\nu}$ at 5~GHz. Column (6): logarithm of 2--10~keV luminosity. Column (7): X-ray weakness factor (see Section \ref{subsec:data_lum}). Column (8): logarithm of $\Rx = \lrlx$. For objects with $\fweak \geq 6$, $\Rx$ has been corrected for X-ray weakness (see Section \ref{subsec:data_lum}). Column (9): X-ray observation source (C = Chandra, X = XMM-Newton, S = Swift, E = eROSITA). Column (10): X-ray data reference. References: B09: \citet{Bianchi09};  CSC: Chandra Source Catalog 2.1 \citep{Evans24}; D09: \citet{Desroches09}; D12: \citet{Dong12a}; ER: eRASS \citep{erass2}; G14: \citet{Gultekin14}; H07: \citet{Holczer07}; P05: \citet{Piconcelli05}; Ri17: \citet{Ricci17}; S10: \citet{Shu10}; W13: \citet{Wang13}.\\
$^{*}$\,For the M93 subsample, the listed VLA flux density ratio is instead A-configuration to D-configuration peak flux density (see Section \ref{subsec:samp3}).
}
\end{deluxetable*}

\section{\texorpdfstring{Notes on Deriving an Expression for $\Lx$}{Notes on Deriving an Expression for the X-ray Luminosity}} \label{sec:app} 

We performed the following derivation as validation for the expression given in Eq.~(\ref{eq:Lx_saturated_gen}).

\citet{Merloni03a} gives the following equation for $f = \Lx/\Lbol$, the fraction of disk power (assuming a radiation pressure-dominated accretion flow) dissipated in the corona:
\begin{equation} \label{eq:f_clos1}
\frac{4\alpha^2_0 - f^4}{f^4(1-f)^2} = 7.3 \times 10^5(\alpha_0 m)^{2/9} r^{-7/3}[\mdot J(r)]^{16/9}.
\end{equation}
In the above, $\alpha_0 = \beta^{-1}(P_{\rm tot}/P_{\rm gas})^{1/2}$ where $\beta = P_{\rm tot}/P_{\rm mag}$ is the plasma parameter, $m = \mbh/\msun$ is the dimensionless mass, $r = R/R_{\rm s} = Rc^2/2G\mbh$ is the dimensionless, scaled radial distance, and $J(r) = 1 - \sqrt{r_{\rm in}/r}$.

To simplify and find an expression for $f$, we assume $f \ll 1$ and $4\alpha^2_0 \gg f^4$ (for pressures near equipartition), approximating the left-hand side of Eq.~(\ref{eq:f_clos1}) via
\begin{equation}
\frac{4\alpha^2_0 - f^4}{f^4(1-f)^2} \approx \frac{4\alpha^2_0}{f^4}.
\end{equation}
We also take $r$ as fixed and ignore its $\mbh^{-1}$ dependence, assuming most radiation is emitted at the same scaled radial distance. These simplifications allow us to rewrite the expression as
\begin{equation} \label{eq:f_LxLbol}
f = \Lx/\Lbol \propto \mbh^{-1/18} \mdot^{-4/9},
\end{equation}
as also noted by, e.g., \citet{Wang13}. Qualitatively, this expectation of relatively weaker X-ray emission at higher $\ledd$ is supported by studies indicating that values of X-ray bolometric correction factors tend to increase with accretion rate \citep[e.g.,][]{Vasudevan09, Lusso12} and that $\aox$ is anticorrelated with $\ledd$ \citep[e.g.,][]{Maithil24}.

We can then incorporate Eq.~(8) of \citet{Mineshige00} to solve for $\Lbol$ (noting that $\Ledd \propto \mbh$) and find:
\begin{equation} \label{eq:Lx_propto}
\Lx =
\begin{cases}
(K_{\rm x}/25\eta) \mbh^{17/18} \mdot^{5/9} & : \frac{\mdot}{\eta} < 50\\
2K_{\rm x} \mbh^{17/18} \mdot^{-4/9} A(\mdot) & : \frac{\mdot}{\eta} \geq 50,
\end{cases}
\end{equation}
where $A(\mdot) = 1 + \ln(\frac{\mdot}{50 \eta})$ and $K_{\rm x}$ is a normalization constant. We note that this expression for $\Lx$ is quite similar to the more generalized case covered by our Eq.~(\ref{eq:FP_LX}) for $q = 0.5$ (which is roughly consistent with empirical trends; e.g., \citealt{Wang04}).

\bibliography{mainref.bib}{}
\bibliographystyle{aasjournalv7}

\end{document}